\documentclass[aps,showpacs]{revtex4-2}
\usepackage{amsmath}    
\usepackage{graphicx}   

\newcommand{\cN}{\mathcal{N}}
\newcommand{\cA}{\mathcal{A}}
\newcommand{\cD}{\mathcal{D}}
\renewcommand{\d}{\mathrm{d}}

\begin{document}
\title{Power-laws in phylogenetic trees and the preferential coalescent}

\author{Stephan Kleinb\"olting$^a$, Nigel Goldenfeld$^b$ and Johannes Berg$^a$}
\email{skleinbo@uni-koeln.de, nigelg@ucsd.edu, bergj@uni-koeln.de}

\affiliation{a University of Cologne, Institute for Biological Physics\\Z\"{u}lpicher Stra{\ss}e 77, 50937 K\"{o}ln, Germany}
\affiliation{b Department of Physics,  
University of California,\\San Diego, 9500 Gilman Drive,
La Jolla, CA 92093, USA}

\begin{abstract}
  Phylogenetic trees capture evolutionary relationships among species and reflect the forces that shaped them. While many studies rely on branch length information, the topology of phylogenetic trees—--particularly their degree of imbalance—--offers a robust framework for inferring evolutionary dynamics when timing data is uncertain. Classical metrics, such as the Colless and Sackin indices, quantify tree imbalance and have been extensively used to characterize phylogenies. 

Empirical phylogenies typically show intermediate imbalance, falling between perfectly balanced and highly skewed trees. This regime is marked by a power-law relationship between subtree sizes and their cumulative sizes, governed by a characteristic exponent. Although a recent niche-size model replicates this scaling, its mathematical origin and the exponent's value remained unclear.

We present a generative model inspired by Kingman’s coalescent that incorporates niche-like dynamics through preferential node coalescence. This process maps to Smoluchowski’s coagulation kinetics and is described by a generalized Smoluchowski equation. Our model produces imbalanced trees with power-law exponents matching empirical and numerical observations, revealing the mathematical basis of observed scaling laws and offering new tools to interpret tree imbalance in evolutionary contexts.
  \end{abstract}

\maketitle

Phylogenetic trees represent the evolutionary relationships between extant organisms. Beyond the direct information concerning shared ancestry, phylogenetic trees are shaped by particular evolutionary forces, which are encoded in the trees and in principle can be inferred from them.  By comparing empirical trees with trees arising from evolutionary models it is possible to infer aspects of the past evolutionary dynamics~\cite{nee1994reconstructed,stadler2013recovering}. 

\begin{figure}[bt]
  \includegraphics[width = .5\textwidth]{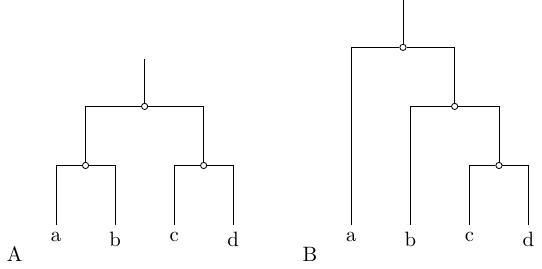}
\caption{{\bf Balance and imbalance in tree topologies.} Both A and B show trees whose nodes all have two offspring (children). 
A shows a perfectly balanced tree, where the total number of descendants of two offspring of a node are always the same. B shows a imbalanced tree, where two offspring of a node have different numbers of descendants (size of the subtree below a node). Tree imbalance affects the distances from the leaves to the root; the distances are same for all leaves in the balanced tree A, but differ across the leaves in unbalanced trees like B. 
\label{fig:treeexample}
}
\end{figure}

In many applications, the times of bifurcation along a phylogenetic tree are uncertain, so only features that do not depend on branch lengths --- the topology of the trees --- can be reliably used.
A widely used family of metrics to analyze and compare trees topologies is based on the imbalance of trees. These metrics quantify if two child nodes bifurcating from an internal node of a tree generally evolve in the same way (leading to a balanced tree), or if they are generally subject to different dynamics (leading to an imbalanced tree). Examples of a balanced tree and an imbalanced tree are are shown in Figure~\ref{fig:treeexample}A and B. For instance, if child nodes generally have very different population sizes, they might undergo further bifurcations at different rates. 
 Several metrics have been developed to quantify the imbalance in phylogenetic trees and are used widely in phylogenetic analysis. The oldest such metric is the Colless index $\mathcal{C}(T)$~\cite{shao1990tree} of a tree $T$. It is given by the total tree imbalance
\begin{equation}
\mathcal{C}(T)= \sum_i \left| A_{i_1} - A_{i_2} \right| \ ,
\end{equation}
where the sum is over all nodes $i$ of the tree apart from leaf nodes, $A_i$ is the size of the subtree defined by $i$ (number of nodes below $i$ including $i$ itself, the size of the so-called clade defined by $i$), and $i_1$ and $i_2$ denote the offspring (children) of $i$. 
The Sackin index~\cite{sokal1983phylogenetic} is defined by the distances of leaf nodes from the root summed over all leaves. Balanced trees generate shorter distances than imbalanced ones. For an overview over these and many other tree balance indices, see~\cite{fischertree}. 

The topological properties of phylogenetic trees have also been studied using the tools of statistical physics. Stich and Manrubia~\cite{stich2009topological} have analyzed the relationship between the subtree size $A_i$ and the cumulative subtree size $C_i=\sum_j A_j$~\cite{herrada2008universal,banavar1999size}, where the sum is over all nodes of the subtree of $i$ (including $i$ itself). For a balanced tree, they find $C \sim A \ln A$ across all nodes, for a maximally imbalanced tree $C \sim A^2$. 

What are the corresponding properties found in empirical phylogenetic trees? Herrada \textit{et al.} found that 
across a wide range of phylogenies, an intermediate scaling $C \sim A^{\eta}$ appears over nearly $3$ decades of data from the TreeBase database, with an average of $\eta = 1.44 \pm 0.01$~\cite{herrada2008universal}. 
For individual trees, the exponent varies roughly between $1.4$ and $1.5$ (where the interval specifies the standard deviation around the mean, see Fig.~3A of~\cite{herrada2008universal}). 
A scaling law compatible with these results has also been found in the Greengenes database~\cite{desantis2006greengenes} of 16S rRNA gene sequences~\cite{jeraldo2012computational,goldenfeld2014looking}. 
Hence, empirical phylogenetic trees lie in a regime between being perfectly balanced and maximally imbalanced, and this is reflected in the particular exponent $\eta$. 

To account for the particular value of the scaling exponent, Xue, Liu and Goldenfeld~\cite{xue2020scale} proposed a \emph{niche-model} of phylogenetic tree dynamics, where each node carries an internal degree of freedom, called the niche size. The size of the ecological niche of a species can be thought of as the population size this species can attain in a given environment.  The niche size affects the supply of future mutations, and hence the rate of further speciation events and of bifurcations in the phylogenetic tree. In the niche size model of Xue, Liu and Goldenfeld, niche sizes are inherited by the children of a node, and change by a small random amount at each bifurcation. 
As a result, the rate of bifurcations can increase and decrease along a particular lineage; in particular the bifurcation rate can increase rapidly along a lineage associated with large population sizes. This can lead to `bursts' in the number of bifurcations along a particular lineage. 
Crucially, the niche size has a sticky boundary; a lineage will not bifurcate again once the niche size has reached zero. In numerical simulations, Xue \textit{et al.} found that this niche-model generates trees forwards in time that are neither perfectly balanced nor maximally imbalanced. Instead, they exhibit a power-law scaling with the exponent $\eta =1.501$ (with a $95\%$ confidence interval of $[1.496, 1.505]$). However, the mathematical origin of this scaling law and the specific value of the exponent remained unclear. 

In the following, we consider a different generative model for phylogenetic trees, using a coalescent-based picture.  In this way, we try to work backwards to understand the link between the evolutionary/ecological
dynamics encoded by the niche model and the statistics of the resulting phylogenetic trees.  At the end of the paper, we discuss briefly the relationship between the two approaches. 
 Coalescent theory considers the generation of a phylogenetic tree from the leaves upwards in individual coalescence events~\cite{kingman1982coalescent}. The dynamics thus runs backwards from the present into the evolutionary past. The simplest approach is to pick pairs of nodes uniformly and coalesce one such pair at each step, yielding the so-called Kingman coalescent~\cite{kingman1982coalescent}. To mimick the niche size dynamics within the coalescent dynamics, we assign each node an internal degree of freedom and use a simple iterative rule, preferential coalescence. The mathematical framework to describe the resulting dynamics turns out to be the Smoluchowski coagulation kinetics~\cite{krapivsky2010kinetic}. 

\begin{figure}[tb]
  A\includegraphics[width = .3\textwidth]{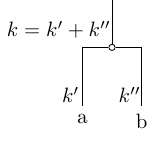}
  \hspace{2cm}
  B\includegraphics[width = .45\textwidth]{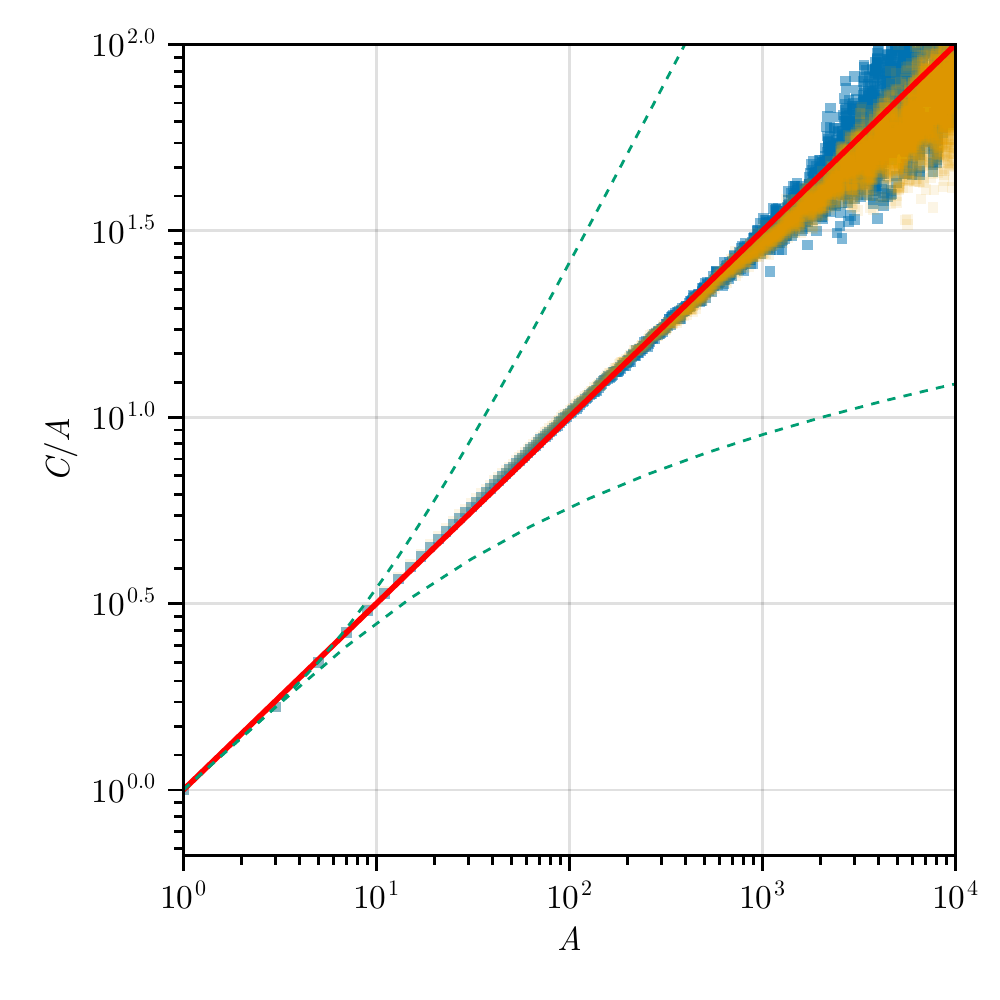}
\caption{{\bf The preferential coalescent and the niche model.} 
A An illustration of the iterative rule~\eqref{eq:massiter} for the niche size at a particular internal node (coalescence event). B A tree of size $N=2^{19}$ was generated using the niche model~\cite{xue2020scale} starting from a single node with niche step size $\sigma=3$. The cumulative subtree size $C$ over the subtree size $A$ is plotted against $A$ (blue symbols) showing the same scaling law as reported in~\cite{xue2020scale}. For comparison, a second tree was generated using the preferential coalescent from $2N$ nodes (orange symbols). 
Dotted lines show the relations between $C$ and $A$ corresponding to perfectly symmetric and perfectly asymmetric trees, respectively (see Figure~\ref{fig:treeexample}). The red line shows the scaling $C \sim A^{3/2}$.
\label{fig:niche_vs_coal}}
\end{figure}

In order to characterize the statistics of genealogical trees we follow~\cite{herrada2008universal,herrada2008universal,xue2020scale} and characterize each node $i$ of a tree by two quantities related to the subtree it induces. The subtree induced by a node $i$ is defined by the tree descending from that node, with $i$ itself as a root. $A_i$ denotes the number of nodes in that subtree (that is the number of nodes below $i$, including $i$). $D_i$ denotes the sum of all distances between $i$ and the nodes in the subtree induced by $i$. Since $C_i \equiv \sum_j A_j = D_i+A_i$, the cumulative distance is as informative as the cumulative subtree size and we will 
treat the two quantities equivalently. 
The cumulative distance is somewhat easier to interpret and allows to make contact with distance-based tree measures like the Sackin index.  


All leaves $i$ of the tree have $A_i=1$ and $D_i=0$. Going up the tree, one can iteratively compute the corresponding values for all nodes using
\begin{align}
  \label{eq:ADiter}
  A&=A'+A''+1 \\
  D&=D'+A'+D''+A'' \nonumber\ ,
\end{align}
where $A$ denotes the size of a subtree of a node just above two nodes with subtree sizes $A'$ and $A''$, and analogously for $D$, see Figure~\ref{fig:niche_vs_coal}A. 

Next, we define a simple coalescent model of niche dynamics. We assign each node $i$ an internal degree of freedom, the niche size $k_i$. At each coalescence event, two nodes are chosen independently with a probability proportional to their niche sizes. This defines preferential coalescence. The newly created parent of these two nodes is assigned the sum of the niche sizes
\begin{equation}
\label{eq:massiter}
k=k'+k'' \ , 
\end{equation}
see Fig.~\ref{fig:niche_vs_coal}A. 
Even when starting from identical niche sizes, preferential coalescence introduces heterogeneity into the tree, with some nodes coalescing much faster than others. 

Figure~\ref{fig:niche_vs_coal}B shows $C/A$ versus $A$ for both the niche model of Xue, Liu and Goldenfeld~\cite{xue2020scale} and the preferential coalescence model. Both are well described by a scaling law $C/A \sim A^{1/2}$ over several decades. This means preferential coalescence produces imbalanced trees and a non-trivial relationship between the observables $A$ and $C$ (or $A$ and $D$). 

To analyse the preferential coalescence dynamics, we note that the dynamics of the niche-sizes defined by equation~\eqref{eq:massiter} maps onto the well-known 
Smoluchowski coagulation kinetics~\cite{smoluchowski1916drei}. 
Under the Smoluchowski model of particle dynamics, pairs of particles collide and form a composite particle, whose mass is equal to the sum of the masses of the two constituent particles. Specifically, coagulation of particles of masses $k'$ and $k''$ occurs at a rate $K_{k'k''}$. The dynamics of the number $n_k(t)$ of particles with mass $k$ is given by the Smoluchowski equation 
\begin{equation}
  \label{eq:smoluchowski}
\partial_t n_k = \frac{1}{2} \sum_{k'+k''=k} K_{k'k''} n_{k'} n_{k''} - n_k \sum_{k'} K_{kk'} n_{k'} \ .
\end{equation}
This is a rate equation (mean-field equation), which 
describes the dynamics of the expectation value of the number $n_k$ of particles of mass $k$. 
The first term of \eqref{eq:smoluchowski} describes the emergence of a new composite particle from particles of masses $k'$ and $k''$. The constraint in the sum implements
\eqref{eq:massiter}. The factor of $1/2$ avoids overcounting the pairs of particles. The second term describes the corresponding loss of particles due to coalescence. Summing \eqref{eq:smoluchowski} over 
the masses $k$ shows that the total mass $M=\sum k n_k(t)$ is a constant of the motion. 
The Smoluchowski equation \eqref{eq:smoluchowski} has been researched extensively in non-equilibrium physics~\cite{krapivsky2010kinetic} and mathematical physics~\cite{aldous1999deterministic}. Applications range from the clustering of nanoparticles~\cite{alexandrov2022dynamics} to swarm size dynamics~\cite{niwa1998school}. Different kernels $K_{ij}$ have been considered~\cite{krapivsky2010kinetic}; $K_{ij}=ij$ defines the preferential aggregation model (also termed multiplicative coalescence or preferential coalescence), where particles coalesce at rates proportional to their masses. 

With the kernel $K_{ij}=ij$, the Smoluchowski equation \eqref{eq:smoluchowski} also describes the dynamics of niche sizes under a coalescent dynamics where nodes are picked for coalescence with a probability proportional to their niche sizes. We now construct analogous equations for the dynamics of the expectations values of $A$ and $D$ iteratively defined by \eqref{eq:ADiter} in order to describe the statistics of these tree observables. We denote the value of $A$ summed over all nodes with niche size $k$ by $a_k$ and the value of $D$ summed over all nodes with niche size $k$ by $d_k$. The dynamics \eqref{eq:ADiter} yields the following rate equations
\begin{align}
  \label{eq:smoluchowskiAD_A}
\partial_t a_k &= \frac{1}{2} \sum_{k'+k''=k} K_{k'k''} \left[a_{k'} n_{k''} + n_{k'} a_{k''} + n_{k'} n_{k''}\right] - a_k \sum_{k'} K_{kk'} n_{k'} \\
\partial_t d_k &= \frac{1}{2} \sum_{i+j=k} K_{ij} \left[(d_{k'}+a_{k'}) n_{k''} + n_{k'} (d_{k''}+a_{k''}) \right] - (d_k+a_k) \sum_{k'} K_{kk'} n_{k'} 
\label{eq:smoluchowskiAD_D}\ .
\end{align}
These equations follow from a generalized Smoluchowski rate equation for the number of nodes with specific values of $k$, $A$, and $D$, 
\begin{equation}
  \label{eq:smoluchowski_nkAD}
\partial_t n_{kAD} = \frac{1}{2} \sum_{\substack{k'+k''=k \\A'+A''+1=A \\D'+A'+D''+A''=D  }}
K_{k'k''} n_{k'A'D'}n_{k''A''D''} - n_{kAD} \sum_{k'} K_{kk'} n_{k'A'D'} \ ,
\end{equation}
where the constraints to the sums over $k',k'',A',A'',D',D''$ implement the iterative rules \eqref{eq:ADiter} and \eqref{eq:massiter}. 
Forming moments $a_k(t) \equiv \sum_{A,D} A n_{kAD}(t)$ and $d_k(t) \equiv \sum_{A,D} D n_{kAD}(t)$ and exploiting these constraints in sums over $A',A'',D',D''$ yields the dynamics of moments \eqref{eq:smoluchowskiAD_A} and \eqref{eq:smoluchowskiAD_D}. The generalized Smoluchowski equation~\eqref{eq:smoluchowski_nkAD} can be defined for any metric on a tree that can be computed recursively. 

The three dynamical equations \eqref{eq:smoluchowski} to \eqref{eq:smoluchowskiAD_D} for $n_k, a_k$, and $d_k$ can be solved using the standard generating function approach for the Smoluchowski equation~\cite{krapivsky2010kinetic}. We define
\begin{align}
  \label{eq:def_generatingfunctions}
\cN(z,t) &=\sum_k k e^{zk} n_k(t) \\
\cA(z,t) &=\sum_k k e^{zk} a_k(t) \\
\cD(z,t) &=\sum_k k e^{zk} d_k(t) 
\end{align}
which yields 
\begin{align}
  \partial_t\cN &=(\cN-M)\partial_z \cN \label{eq:partialcN}\\
  \partial_t\cA &=(\cA+\cN)\partial_z \cN + (\cN-M)\partial_z \cA \label{eq:partialcA}\\
  \partial_t\cD &=(\cD+\cA)\partial_z \cN + \cN\partial_z \cA + (\cN-M)\partial_z \cD \label{eq:partialcD}\ .
  \end{align}
These partial differential equations can be solved using the method of characteristics. Crucially, 
the characteristics for the three equations are the same: (i) the terms depending on the partial derivatives of 
$\cN,\cA,\cD$ w.r.t. time $t$ on the left hand side of each equation have the same coefficients (unity), and (ii) the terms depending on the partial derivatives of $\cN,\cA,\cD$ w.r.t. $z$ on the left hand side also have the same coefficients $\cN-M$ in the corresponding equations. Because of (i), the independent variables of the characteristics are all equal to $t$ (up to constants set to zero) and will not be assigned a new symbol. Due to (ii), $\frac{\d z}{\d t}= -(\cN-M)$. The latter is solved subject to the initial condition $\cN(z,t=0)=M e^z$ (starting with equal niches sizes equal to one) by 
\begin{equation}
  \label{eq:z_sol}
  z=-(\cN-M)t + \ln\left(\cN/M \right) \ .
\end{equation}
The three ordinary differential equations for $\cN,\cA,\cD$ resulting from \eqref{eq:partialcN} to \eqref{eq:partialcD}
are
\begin{align}
\frac{\d \cN}{\d t} &= 0 \\
\frac{\d \cA}{\d t} &= (\cA+\cN) \partial_z \cN \label{eq:ordinarydiff_A}\\
\frac{\d \cD}{\d t} &= (\cD+\cA) \partial_z \cN + \cN\partial_z \cA  \label{eq:ordinarydiff_D}\ .
\end{align}
The generating function $\cN$ is thus constant along the characteristics. The dynamics of $\cA$ along the characteristics will be solved based on $\partial_z \cN$ along the characteristic (calculated 
from the solution of $\cN$), and likewise for the dynamics of $\cD$. 

The solution for $z(t,N)$ \eqref{eq:z_sol} allows to solve the Smoluchowski equation~\eqref{eq:smoluchowski} by determining the coefficients of the generating function $\cN(z,t)$ using the Lagrange inversion formula~\cite{krapivsky2010kinetic}: Re-arranging \eqref{eq:z_sol} gives
\begin{equation}
  \label{eq:z_sol2}
  \cN t e^{-\cN t}= Mt e^{z-Mt}\ .
\end{equation}
Writing $x \equiv Mt e^{z-Mt}$ and $y \equiv \cN t$ the resulting equation $y e^{-y}=x$ is solved by $y(x)=\sum_k y_k x^k$, where 
\begin{equation}
y_k=\frac{1}{2 \pi i} \oint \d x y(x)/x^{k+1} = k^{k-1}/k! \ ,
\end{equation}
where the last step involved a change of integration variable to $y(x)$~\cite{krapivsky2010kinetic},
and hence from \eqref{eq:def_generatingfunctions}
\begin{equation}
  n_k= \frac{k^{k-2}}{k!} (Mt)^k/t e^{-Mtk} \approx \frac{1}{\sqrt{2 \pi}\, t} k^{-5/2}(Mt)^k e^{-k(Mt-1)}\ ,
  \end{equation}
where Stirling's formula was used in the last step.

  The same strategy can be used to calculate the coefficients $a_k(t)$ and $d_k(t)$: Differentiating equation~\eqref{eq:z_sol} linking $z$ and $\cN$ yields $\partial_z \cN =\cN/(1-t \cN)$. This allows to solve the ordinary differential equation for $\cA$ \eqref{eq:ordinarydiff_A}, giving $t \cA=\frac{1+t \cN}{1-t \cN}$. Expressing $t \cN$ as a function of $t \cA$ in \eqref{eq:z_sol2} allows again to extract the coefficients of the generating function, giving 
  \begin{equation}
    \label{eq:ak_final}
    a_k= \frac{k^{k-2}(2k-1)}{k!} (Mt)^k/t  e^{-Mtk} \approx \frac{2}{\sqrt{\pi}\, t} k^{-3/2}(Mt)^k e^{-k(Mt-1)}\ .
    \end{equation}
    The same strategy applied to $\cD$ gives 
    \begin{equation}
      t \cD = \frac{2 (t\cN)^3-2 t\cN(1-t\cN)\ln(1-t\cN)}{(1-t\cN)^2} \ .
      \end{equation}
      With the shorthands $r=t\cD=\sum_k r_n x^n$ and $x \equiv Mt e^{z-Mt}$ as above, expressing $y\equiv t \cN$ as a function of $t \cD$ gives 
      \begin{equation}
r_k=\frac{1}{2 \pi i} \oint \d y \left[ \frac{2 y^3}{1-y} - 2 y \ln(1-y) \right] \frac{e^{ky}}{y^{k+1}} \approx e^k
      \end{equation}
      and thus 
      \begin{equation}
        \label{eq:dk_final}
        d_k \approx  (Mt)^k/t e^{k-Mtk} k^{-1}\ .
        \end{equation} 
The results for $n_k,a_k,d_k$ show different power-laws as well as a cutoff resulting from the term 
$(Mt)^k e^{k-Mtk} = \exp\{ -k\left[Mt-1-\ln(Mt)\right] \}$ (which can be neglected when $k$ much less than  
$\left[Mt-1-\ln(Mt)\right]^{-1}$). These power-laws are the well-known $n_k \sim k^{-5/2}$ of the Smoluchowski equation for preferential aggregation as well as power-laws describing coalescent trees under preferential coalescence with  
$a_k \sim k^{-3/2}$ and $d_k \sim k^{-1}$.  

To compare these results to the coalescent, we note that the time-scale defined by the Smoluchowski equation~\eqref{eq:smoluchowski} is not measured in aggregation steps. Summing over $k$ in \eqref{eq:smoluchowski} shows that the total number of particles (or nodes in the coalescent) obeys $\partial_t \sum_k n_k = M^2/2$, which means the number of aggregation steps (on average) is $M^2 t/2$. 

\begin{figure}[tbh]
          A
          \includegraphics[width = .3\textwidth]{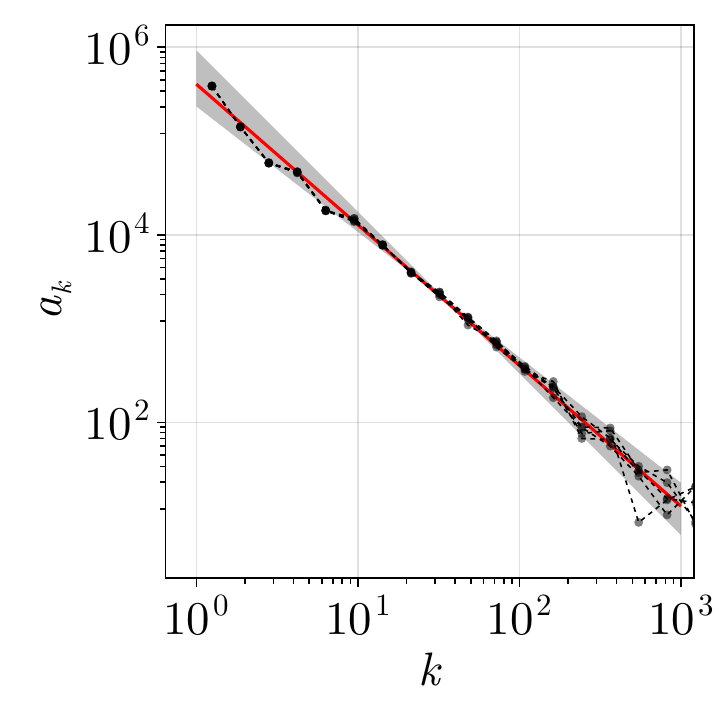}
          B
        \includegraphics[width = .3\textwidth]{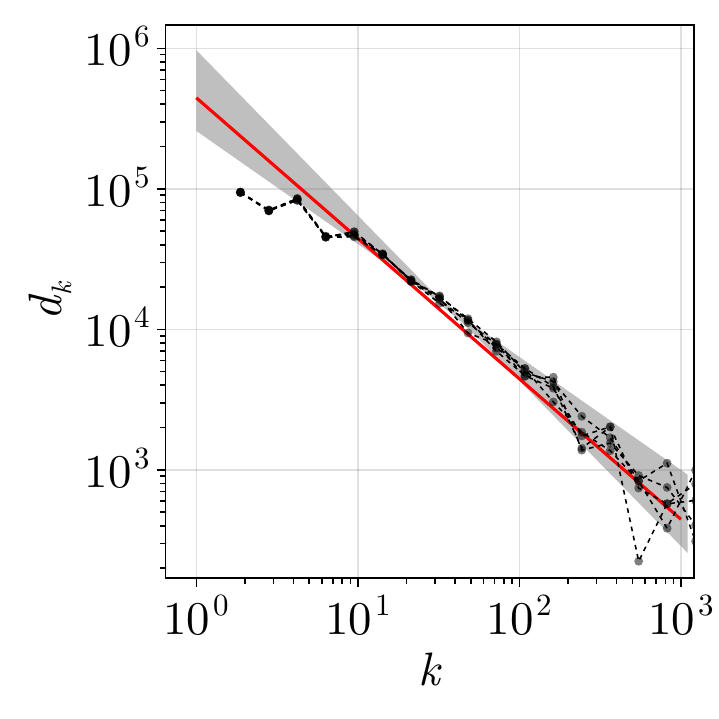}
        C
        \includegraphics[width = .3\textwidth]{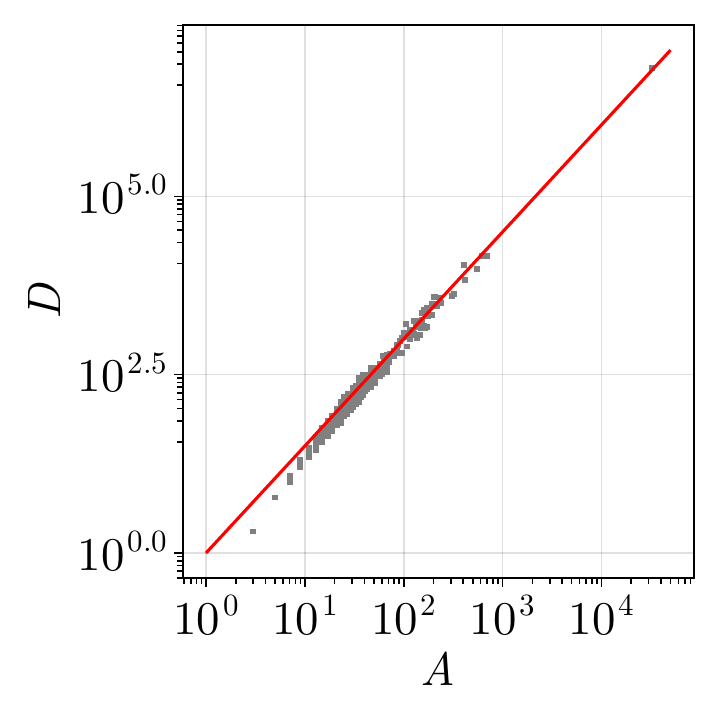}
        \caption{{\bf Preferential coalescence and the balance of genealogical trees.}
        We compare the results of numerical simulations with the power laws \eqref{eq:ak_final} and \eqref{eq:dk_final} derived for the preferential coalescence model \eqref{eq:massiter} and \eqref{eq:ADiter}. For the simulations, we start the coalescence process with $N=2^{19}$ nodes of niche size $k=1$, $A=1$ and $D=0$ and coalesce nodes according to the preferential aggregation rule and the iterative rules \eqref{eq:massiter} and \eqref{eq:ADiter} until the number of nodes has halved. We show the statistics of $A$ and $D$ and $k$ for the remaining nodes. 
        \textbf{A} To compute $a_k$ from simulations, the sizes $A$ of the subtrees below nodes with niche sizes $k$ are summed up, followed by a logarithmic binning on intervals $[1.5^n, 1.5^{n+1})$. A scatter plot of multiple runs along with the power-law $a_k \sim k^{-3/2}$ \eqref{eq:ak_final} in red is shown on a doubly-logarithmic scale. A regression line is fitted for each simulation. 
        The grey banded region is the envelope of regression lines resulting from 100 independently simulated trees.
        \textbf{B} The results of the same procedure repeated with the cumulative distance $D$ compared to the power-law $d_k \sim k^{-1}$ resulting from  
        \eqref{eq:dk_final}. 
        \textbf{C} The cumulative distances $D$ are shown against subtree size $A$ across all remaining nodes. 
        The straight line shows the relation $D \sim A^{3/2}$ which follows from the power-laws \eqref{eq:ak_final} and \eqref{eq:dk_final}. 
        \label{fig:powerlaws}}
        \end{figure}

In Figure~\ref{fig:powerlaws}, we compare our results to numerical simulations. A single phylogenetic tree is built under preferential coalescence, starting with $N=2^19$ nodes each of niche size $k=1$. 
We run the coalescence process until half the original number of nodes is left and compute the values of $A$ and $D$ for the remaining nodes. (This stopping criterion was used since it is the point where since $M^2t/2=M/2$ we have $Mt=1$ and no exponential correction to the power-laws.)
Fig.~\ref{fig:powerlaws}A and B show $a_k$ and $d_k$, respectively against $k$, along with the power-laws with exponents $-3/2$ and $-1$ (red lines). Fig.~\ref{fig:powerlaws}C shows $D$ versus $A$ for all nodes of the tree. If the variation of these variables is due to variation in $k$, \eqref{eq:ak_final} and \eqref{eq:dk_final} predict $D \sim A^{3/2}$. These power laws agree with 
the results of numerical simulations shown in Fig.~\ref{fig:powerlaws}. The scaling law $D \sim A^{3/2}$ also agrees very well with the simulations reportedin~\cite{xue2020scale} and the scaling $C/A \sim A^{1/2}$, and may also account for the scaling seen empirically in~\cite{herrada2008universal}. 

We note that the niche size dynamics of Xue, Liu and Goldenfeld~\cite{xue2020scale} and the model analyzed here are very different, yet they produce the same scaling exponents. In~\cite{xue2020scale}, time runs forwards, and the niche size of a species changes by random increments along a lineage. An increment of the niche size at a particular internal node $i$ increases the probability of speciation, thus enhancing the size of the subtree $A_i$. Small niche-sizes tend to have small subtrees, and, via a sticky boundary (niche-size zero) tend to remain at small niche-sizes. This generates heterogeneity in the tree generated by time running forwards.
By contrast, in our coalescent-based analysis time runs backwards and 
the heterogeneity is generated by the preferential coalescence rule. This rule makes nodes with large niche sizes coalesce faster and increase their niche-size further. Despite the very different origins of tree heterogeneity in these two mechanisms the resulting scaling laws are the same, pointing to a mechanism of universality that remains to be understood. 

Understanding this link between a forward-time birth-death model a coalescent model presents a significant challenge. 
The reason is that under a forward-time birth-death dynamics, birth and death events are independent of each other (given growth rates or niche sizes). As a result, times between these events are exponentially distributed. On the other hand, when considering time running
backwards, events are correlated by the non-trivial conditions induced by the reconstruction of the phylogenetic tree. As a result, times between events do not follow simple exponential distributions. For instance, the distribution of branch lengths in a phylogenetic tree are not exponential, because of the conditions that the clades established at the beginning and at the end of a branch do not die out, and birth events along the branch do not establish additional extant clades~\cite{stadler2012distribution,dieselhorst2024branch}. 
A coalescent model ignores these conditions, leading to a tractable model where nodes coalesce at some rate. This rate can depend on population size (like in Kingman's coalescent model~\cite{kingman1982coalescent}), or on the niche size in the preferential aggregation model considered here. Why this simplification of the statistics of the reconstructed phylogenetic tree yields correct exponents remains an open question. 

{\bf Acknowledgements:} 
	This work was funded by the Deutsche Forschungsgemeinschaft (DFG, German Research Foundation) grant SFB1310/2 - 325931972. Many thanks to Joachim Krug for discussions.

\bibliographystyle{abbrvnat}
\bibliography{prefagg}

\end{document}